\documentstyle [12pt]{article}
\title{
\hspace{3.0truein}{\small IFT-494-UNC}\\
\vspace{0.2truein}
{Thermodynamics of an Anyon System}
}
\author{Wei Chen\footnotemark[1] \\
Department of Physics\\
University of North Carolina\\
Chapel Hill, NC 27599-3255}
\date{ Physical Review D (in press)  }
\footnotetext{$^*$ chen@physics.unc.edu}

\begin{document}
\maketitle
\vspace{0.2truein}
\begin{abstract}
We examine the thermal behavior of a relativistic
anyon system, dynamically realized by coupling a charged massive
spin-1 field to a Chern-Simons gauge field. We calculate the
 free energy (to the next leading order), from which all
thermodynamic quantities can be determined. As examples,
the dependence of particle density on the anyon statistics and
the anyon anti-anyon interference
in the ideal gas are exhibited. We also calculate two and three-point
correlation functions, and uncover
certain physical features of the system in thermal equilibrium.
\end{abstract}

\newpage

\renewcommand{\theequation}{\thesection.\arabic{equation}}
\baselineskip=18.0truept
\vskip 0.5truein
\section{Introduction}
\setcounter{equation}{0}
\vspace{3 pt}

Statistics of particles plays a fundamental role in determining
macroscopic properties of many body systems. The conventional particles
are classified into bosons and fermions as they obey either
Bose-Einstain or Fermi-Dirac statistics. It is known that a many-body
wave-function is symmetric under permutations of identical bosons, but
it is anti-symmetric for identical fermions. And,
bosons condense while fermions exclude.
Attempts to generalize statistics date back at least to Green's
work in 1953 \cite{Green}.
Green found that the principles of quantum mechanics also allow two kinds of
statistics called parabose statistics and parafermi statistics.
Another type of interpolating statistics is provided by the concept of anyons
\cite{any}\cite{Wil}. Limited in two spatial dimensions,
anyons are particles (or excitations)
whose wave-functions acquire an arbitrary
phase, $e^{i\pi\alpha}$, when two of them are braided.
The phase factor $\alpha$, now being any value between 0 and 1 (modular 2),
defines the fractional statistics of anyons.
The concept of fractional statistics (or anyons) has been useful in
the study of certain important condensed
matter systems, particularly in the theories of quantum Hall effect \cite{L},
superconductivity \cite{L1} and some other strongly
correlated systems \cite{LRF}. Most efforts to understand the
fractional statistics have been in absolute zero, though
in literature one can also find calculations for certain quantities
at finite temperatures such as corrections to the statistical
parameter and induced matter masses \cite{tem}. Systematic
study to understand thermodynamics of anyon systems
should be necessary. In this Paper we will address the issue.

An elegant dynamical construction of particle system that obeys
fractional statistics is to couple bosons or fermions via a conserved
current to a Chern-Simons gauge field, so that the  fictitiously charged
particles each is endowed with a ``magnetic'' flux.
The flux carrying charged particles are nothing but anyons,
as  when one such particle winds around another,
 it acquires indeed a Aharonov-Bohm phase \cite{Wil}.
Since the Chern-Simons coupling characterizes
the strength of the attached flux and thus the Aharonov-Bohm phase,
it characterizes the fractional statistics of the particles as well
and is called the statistical parameter.
This sort of Chern-Simons constructions involves only local interactions,
and so is readily to be dealt with at the level of local quantum
field theory.  In study of some systems that are basically non-relativistic,
low energy effective theories are found sufficient, convenient, or both.
On the other hand, however, to understand the short distance behavior
of a system, including anyon anti-anyon pair
productions, a relativistic treatment is necessary.
In the relativistic case, one has found that a
theory with a scalar minimaly coupled to a Chern-Simons field at a
particular value of Chern-Simons coupling is equivalent to the
theory for free (spin-1/2) Dirac fermions \cite{P}.
This phenomenon is called statistics
transmutation. In fact, here, not only the statistics of
the matter field is transmuted, but also the spin of it.
It is further demonstrated that in
a spinning matter Chern-Simons field theory
with an arbitrary Chern-Simons coupling, an integer
(or odd-half-integer) part of the Chern-Simons coupling can
be reabsorbed by changing the spin, the character of Lorentz
representation, of the spinning matter field  \cite{CI}.
Once this is done, in the resulting theory the matter field has a
higher spin and the Chern-Simons coupling is weaker.
This implies there exist many equivalent
field theory representations for one single anyon system.

In this Paper, we consider thermodynamics of a free relativistic
anyon system described by a massive spin-1 field coupled to the
Chern-Simons gauge field. Using the finite temperature field
theory method \cite{K}, we calculate the free-energy, from which
all thermodynamical quantities can be obtained.
In particular, we exhibit the particle density as a function of the
fractional statistics, and the interference between anyons and anti-anyons
in the ideal gas.  We also calculate the two- and three-point correlation
functions to obtain certain physical quantities  such as the screening length,
effective masses and temperature dependent statistical parameter.
Reliability of perturbative expansion requires
the Chern-Simons coupling being small. Under this restriction,
recall the spin and statistics transmutation \cite{CI},
this theory of free anyons is equivalent to a theory of
charged  spin-1/2 particles (electrons for instance)
each particle carries about one unit of flux.
The perturbative results obtained here could be directly useful
only to this system and alike. For a case in which an electron
carries more flux, one may map the problem to
a theory with a higher spin matter field and weaker Chern-Simons coupling
and conduct a similar perturbative calculation.

\renewcommand{\theequation}{\thesection.\arabic{equation}}
\vskip 0.5truein
\section{
The Model}
\setcounter{equation}{0}
\vspace{3 pt}

The model of interest is given by
\begin{equation}
I = \int_\Omega
\frac{1}{2}\left(
\epsilon_{\mu\nu\lambda}
B^*_\mu(\partial_\nu - i2ga_\nu)B_\lambda
+MB^*_\mu B_\mu
+ \epsilon_{\mu\nu\lambda}
a_\mu\partial_\nu a_\lambda\right)\;,
\label{S}
\end{equation}
where the three space-time manifold $\Omega$ has a Lorentzian signature.
First of all, a free massive spin-1 $B_\mu$ theory, something like
Eq.(\ref{S}) with a real $B_\mu$ and without
the Chern-Simons interaction, was first proposed as a self-dual
field theory \cite{TPN}. This self-dual theory, possessing
a single massive degree of freedom and governed by a first
derivative order action, was then shown to be
equivalent (by a Legendre transformation)
to topologically massive electrodynamics \cite{DJ}.
It was also shown to have positive Hamiltonian and positively definite norm
of states in Hilbert space \cite{IIM}. The free massive
spin-1 field theory was used as a starting theory to construct
relativistic wave equation for anyons by considering
one-particle states as unitary representations of the
Poincare algebra in 2+1 dimensions \cite{JN}.

Then, we now consider another construction of anyons as shown in
Eq.(\ref{S}) by using a  complex $B_\mu$ field, and coupling it to a
Chern-Simons field via the current
\begin{equation}
j_\mu = -i\epsilon_{\mu\nu\lambda}B^*_\nu B_\lambda\;.
\end{equation}
This current is conserved as Eq.(\ref{S}) is invariant under global
U(1) transformations. As stated in the previous section,
a Chern-Simons coupling endows charged particles
with fluxes and turns them into anyons. To check this, let's
consider the equation of motion of the $a_\mu$ field
\begin{equation}
\epsilon_{\mu\nu\lambda}\partial_\nu a_\lambda =
2gj_\mu\;,
\label{eq1}
\end{equation}
in particular, $gj_0=b=\frac{1}{2}\epsilon_{ij}\partial_ia_j$
for $\mu=0$. This implies a charged particle with density $j_0$
is attached with a magnetic flux tube $b$. 
The parameter $g$ characterizes the combined strength of charge
and flux, and thus the statistics of anyons. Eq.(\ref{eq1}) also implies that
Chern-Simons field has no independent dynamical degree
of freedom. Indeed, the equation
of $a_\mu$ could be solved by integrating over the current.
Equivalently, one could integrate out the $a_\mu$ field
from the action Eq.(\ref{S}) and obtain a non-local
term for $B_\mu$ (which we are not going to do in this work).

Eq.(\ref{S}) is also invariant under local U(1)
gauge transformations:
\begin{eqnarray}
a_\mu ({\bf x},t) &\rightarrow& a_\mu ({\bf x},t)
+ \partial_\mu \alpha ({\bf x},t)\;,\label{t1}\\
B_\mu ({\bf x},t) &\rightarrow& e^{i\alpha({\bf x},t)}B_\mu({\bf x},t)\;.
\label{t2}
\end{eqnarray}
Namely, $a_\mu$, as a gauge field, fills adjoint representation of
the gauge group while $B_\mu$, like a charged matter field, fills
fundamental representation, though both $a_\mu$ and $B_\mu$ are governed
by a first derivative order kinetic term.
The non-zero mass $M$ of the $B_\mu$ field plays a key role here. It is
the mass that makes $B_\mu$ a (matter) field that carries
local dynamical degrees of freedom. Indeed,
if setting $M=0$ in Eq.(\ref{S}), one obtains
a topological Chern-Simons  $SU(2)$ gauge theory \cite{CI}.

To see how many degrees of freedom  the massive
spin-1 field $B_\mu$ carries, 
let's write down the equation of motion for $B_\mu$
\begin{equation}
\epsilon_{\mu\nu\lambda}(\partial_\nu-i2ga_\nu)B_\lambda + i MB_\mu = 0\;.
\label{EQ}
\end{equation}
Action $(\partial_\mu-i2ga_\mu)$ on Eq.(\ref{EQ}), and using Eq.(\ref{eq1})
and the current conservation $\partial_\mu j_\mu =0$, for $M\neq 0$,
we have
\begin{equation}
(\partial_\mu - i2ga_\mu) B_\mu = 0\;.
\label{cons}
\end{equation}
Eq.(\ref{cons}) is actually a constraint. In canonical approach,
it is convenient to eliminate $B_0$ by solving Eq.(\ref{cons}).
Since the action involves only first derivative, $B_1$ and $B_2$
are actually canonical conjugate one another. This implies that
$B_\mu$ carries a single degree of freedom per read field.
We will verify this argument in a different perspective when the
free-energy is calculated in the next section.

The terms of first derivative in Eq.(\ref{S}) violate parity and time
reversal symmetries, so does Eq.(\ref{S}) itself.
This is one of the attractive features of anyons related to quantum Hall
effect and some other planar systems. However, Eq.(\ref{S}), being
a free anyon theory, is too simple to be realistic. In practical problems,
more ingredients are usually necessary. For instance, to study
optical features of superconductivity, a dynamical electromagnetic
field and a Chern-Simons field, with gauge
symmetry breakings, are introduced \cite{WZ}; to study the dispersion relations
of electromagnetic waves in an anyon model, both topologically massive gauge
field and Chern-Simons gauge field are used \cite{BLL}.
Though we don't expect to go far in applications in this work,
we hope the study of the simpliest model captures certain basic
features of anyons.

Since a local Chern-Simons interaction is introduced to present
even free anyons, we are dealing with an interacting field theory,
with a dimensionless coupling constant $g$.
In quantum field theories, it happens quite often for
a coupling constant to receive non-trivial renormalization.
However, it is not the case for the Chern-Simons coupling $g$,
because of the topological nature of the Chern-Simons term.
The beta function of $g$ vanishes identically,
and so $g$ is not a running coupling constant. Therefore, the
Chern-Simons coupling serves well as a controlling
parameter in a perturbation expansion. In a heat bath,
a coupling turns out to be temperature dependent.
If the Chern-Simons coupling would go up rapidly with temperature,
perturbation broke down very soon. However, as to be seen below,
the effective Chern-Simons coupling is just a slowly increasing
function of temperature, in a large range of temperatures
perturbation expansion should be reliable.

\renewcommand{\theequation}{\thesection.\arabic{equation}}
\vskip 0.5truein
\section{Free Energy and Particle Density}
\setcounter{equation}{0}
\vspace{3 pt}

Now we attach the system to a heat bath at temperature $T$.
As is known \cite{K}, finite temperature behavior of any theory
is specified by the partition function
\begin{equation}
Z={\rm Tr}e^{-\beta (H-\mu N)}\;;
\end{equation}
and the thermal expectations of physical observables
\begin{equation}
<{\cal O}> ~= \frac{1}{Z}
{\rm Tr}[{\cal O}e^{-\beta (H-\mu N)}]\;,
\end{equation}
where $\beta = 1/T$ is the inverse temperature (the Boltzmann constant
$k_B = 1$); $H$ the Hamiltonian; $N$ the particle
number operator; and $\mu$ the chemical potential,
which appears as a Lagrange multiplier when
the system conserves the particle numbers.

{}From functional integral representation
of a quantum field theory, it is readily to work out
the partition function of the anyon system
at finite temperature $T$. The trick is rather simple: to
replace the time variable $t$ with
the imaginary time $i\tau$ via a Wick rotation,
and to explain the final imaginary time
as the inverse temperature $\beta = 1/T$.
Then the partition function of the system described by
Eq.(\ref{S}) is
\begin{equation}
Z = {\cal N}\int\prod_\mu\prod_\nu\prod_\lambda{\cal D}a_\mu
{\cal D}B^*_\nu
{\cal D}B_\lambda
{\cal D}c
{\cal D}\bar{c}
{\rm exp}\left( -\int_0^\beta d\tau\int d^2x
{\cal L}\right)\;,
\label{Z}
\end{equation}
with the Euclidean Lagrangian
\begin{equation}
{\cal L}=
-\frac{i}{2}\epsilon_{\mu\nu\lambda}
B^*_\mu(\partial_\nu - i2ga_\nu+\delta_{\lambda 0}\mu)B_\lambda
+\frac{M}{2}B^*_\mu B_\mu
-\frac{i}{2}\epsilon_{\mu\nu\lambda}
a_\mu\partial_\nu a_\lambda
+ (\partial_\mu\bar{c})(\partial_\mu c)
+\frac{1}{2\rho}(\partial_\mu a_\mu)^2
\;,
\label{Se}
\end{equation}
where the chemical potential $\mu$ is introduced to
reflect the particle conservation, and the Faddeev-Popov ghosts
$c$ and $\bar{c}$ and the last term above are for covariant gauge
fixing.  Being bosons or ghosts with
ghost number $\pm 1$, all fields in Eq.(\ref{Se}) are
subject to periodic boundary condition so that
\begin{equation}
a_\mu(\beta, {\bf x})=a_\mu(0, {\bf x})
{}~~~~ {\rm and} ~~~~B_\mu(\beta, {\bf x})=B_\mu(0, {\bf x})\;.
\label{bc}
\end{equation}

{}From the partition function Eq.(\ref{Z}),
it is easy to work out the Feynman rules at
finite temperature: the Chern-Simons propagator in the
Landau gauge ($\rho = \infty$) and the vertex are
\begin{equation}
D^0_{\mu\nu}(p)
= \frac{\epsilon_{\mu\nu\lambda}p_\lambda}{p^2}\;,
{}~~~~{\rm and} ~~~~
\Gamma^0_{\mu\nu\lambda}=g\epsilon_{\mu\nu\lambda}\;,
\label{R}
\end{equation}
with $p_3 = 2\pi n T$, which is due to the periodic
boundary condition; and the propagator for the $B_\mu$ field
\begin{equation}
{}~G^0_{\mu\nu} = \frac{\epsilon_{\mu\nu\lambda}p_\lambda
+\delta_{\mu\nu}M+p_\mu p_\nu/M}{p^2+M^2}\;,
\label{R1}
\end{equation}
where  $p_3 = 2\pi n T - i\mu$.
Besides, each loop in a Feynman diagram
carries an integration-summation $T\sum_n\int d^2p/(2\pi)^2$
over the internal momentum-frequency $({\bf p},p_3)$; and at each
vertex, momentum-frequency conservation is required.

The single most important function in thermodynamics
is the free energy, from which all thermodynamic properties
are determined. Now we consider the perturbation
expansion of the free energy. With a conventional Fourier
transformation, we choose to work in
the momentum space. At the leading order,
 we ignore the interaction and calculate
\begin{equation}
Z_0 =
[{\rm det}(-\epsilon_{\mu\nu\lambda} p_\lambda + M\delta_{\mu\nu})]^{-1}
[{\rm det}(-\epsilon_{\mu\nu\lambda} p_\lambda
+ \frac{1}{\rho}p_\mu p_\nu)]^{-\frac{1}{2}}
{\rm det}(p^2)\;.
\label{Z0}
\end{equation}
These determinants are the Gaussian integrals for
the free $B_\mu$, $a_\mu$ and $c$ fields, respectively.
The determinant for the massive $B_\mu$ field is
\begin{equation}
[{\rm det}(-\epsilon_{\mu\nu\lambda}p_\lambda
+ M\delta_{\mu\nu})]^{-1} = \frac{1}{\beta M}
\prod_n\prod_{{\bf p}}[\beta^2 (p^2+M^2)]^{-1}\;.
\end{equation}
The determinant for the Chern-Simons field $a_\mu$ is
\begin{equation}
[{\rm det}(-\epsilon_{\mu\nu\lambda} p_\lambda
+ \frac{1}{\rho}p_\mu p_\nu)]^{-\frac{1}{2}}
= \sqrt{\beta\rho}
\prod_n\prod_{{\bf p}}(\beta^2 p^2)^{-1}\;.
\end{equation}
However, this contribution from $a_\mu$ is canceled out by that
from the ghost, the last determinant in Eq.(\ref{Z0}),
upto a gauge parameter term $\sqrt{\beta\rho}$
which can be absorbed into the zero-point energy.
(Therefore it requires a deduction of the zero-point energy
from the free energy before a gauge fixing, for instance
$\rho = \infty$ in our choice.) This result
is compatible with the fact that the Chern-Simons gauge
field carries no local dynamical degree of freedom. Put together all these,
by definition of free energy density $ {\cal F} = -{\rm ln}Z/(\beta V)$,
we obtain
\begin{equation}
 {\cal F}_0 = 
 T\int\frac{d^2p}{(2\pi)^2}\left(
 {\rm ln}(1-e^{-\beta(\omega-\mu)})
+ {\rm ln}(1-e^{-\beta(\omega+\mu)})\right)\;,
\label{F0}
\end{equation}
with $\omega = \sqrt{{\bf p}^2+M^2}$, and with the zero-point energy
droped. Eq.(\ref{F0}) is recognized as the free energy of
two dimensional, relativistic massive boson ideal gas.
The second term in Eq.(\ref{F0}) is due to the anti-particles.
This verifies that $B_\mu$ carries one single
 dynamical degree of freedom per real field.
Given ${\cal F}$, one may calculate all
other thermodynamic quantities by computing
certain derivatives \cite{Huang}. For instance,
the particle density $n =N/V=-\partial {\cal F}/\partial\mu$.
At the leading order,
\begin{equation}
n_0=\int \frac{d^2p}{(2\pi)^2}
\left(\frac{1}{e^{\beta(\omega-\mu)}-1}
-\frac{1}{e^{\beta(\omega+\mu)}-1}\right)
=\frac{T^2}{2\pi}[g_2(x,-r)
-g_2(x,r)]\;,
\end{equation}
 where we have introduced dimensionless parameters
$x = M/T$ and $r=\mu/T$; and
\begin{equation}
g_n(x,r) = \frac{1}{\Gamma(n)}\int_0^\infty dy y^{n-1}
\frac{1}{e^{\sqrt{x^2+y^2}+r} - 1}\;.
\end{equation}
In particular, $g_2(x,r) = (x+r)^2/2+r^2/2 - {\rm ln}(1-e^{-(x+r)})
+{\rm dilog}(e^{x+r})$.

To consider quantum corrections, we first introduce
a formula that maps the discrete summation
$T\sum^{\infty}_{n=-\infty}f(p_3=2\pi Tn -i\mu)$
into continuum integration.
It holds
\begin{equation}
T\sum^{\infty}_{n=-\infty}f(p_3)=
\frac{1}{2}\int_{-\infty}^\infty \frac{dz}{2\pi}[f(z)+f(-z)]
+\int_{-\infty+i\epsilon}^{\infty+i\epsilon}
\frac{dz}{2\pi}\left(\frac{f(z)}{e^{-i\beta z+\beta\mu}-1}
+\frac{f(-z)}{e^{-i\beta z-\beta\mu}-1}\right)\;,
\label{f}
\end{equation}
as long as the function $f(p_3)$ has
no singularity along the real $p_3$ axis.
This formula is also convenient for regularization, as the
temperature independent part of a quantity is completely
separated out, and as is known \cite{Wein},
only the temperature independent part may
contain divergence and so need ultraviolet (or infrared)
regularization and renormalization. For the later use, let us calculate
\begin{equation}
J(M,T,\mu)=T\sum^\infty_{n=-\infty}
\int\frac{d^2p}{(2\pi)^2}\frac{1}{p^2+M^2}\;.
\end{equation}
By using Eq.(\ref{f}), we have
\begin{eqnarray}
J(M,T,\mu)&=& \int_{-\infty+i\epsilon}^{\infty+i\epsilon}
\frac{dz}{2\pi}\int\frac{d^2p}{(2\pi)^2}
\frac{1}{z^2+\omega^2 }\left(\frac{1}{e^{-i\beta z+\beta\mu}-1}
+\frac{1}{e^{-i\beta z-\beta\mu}-1}\right)\nonumber\\
& &+\int\frac{d^3p}{(2\pi)^3}\frac{1}{p^2+M^2}\;,
\label{J}
\end{eqnarray}
where $\omega^2 = {\bf p}^2+M^2$. The second term in Eq.(\ref{J})
is temperature independent, and it is linearly divergent.
Therefore, a regularization is needed.
If a naive cutoff $\Lambda$ is introduced, the result is
\begin{equation}
\int_\Lambda\frac{d^3p}{(2\pi)^3}\frac{1}{p^2+M^2}
=\frac{\Lambda}{2\pi^2} - \frac{M}{4\pi}\;.
\label{ze}
\end{equation}
One can, of course, choose other proper regularization procedures,
for instance  the regularization by dimensional continuation.
Using the latter regularization to calculate the same integral,
one ends up with only the second term in Eq.(\ref{ze}).
Namely that, the two regularization
schemes differ one another by a $\Lambda$-dependent term.
This linear cut-off term can be absorbed by
renormalization -- re-definitions of the zero temperature
mass and/or coupling constant --
as we shall see below. As long as renormalized quantities
are concerned, no physics should be affected
by the regularization procedure(s) used.
The first integration in Eq.(\ref{J})
involves no divergence, neither ultraviolate nor infrared,
thanks to the Bose-Einstein distribution function.
Performing the integrations on the complex
$z$ plane and in the real two-dimensional ${\bf p}$
space in the first term of Eq.(\ref{J}), we obtain
\begin{equation}
J(M,T,\mu) = \frac{\Lambda}{2\pi^2} - \frac{M}{4\pi}
+\frac{T}{4\pi}[h_2(x,-r)
+h_2(x,r)]\;,
\label{J1}
\end{equation}
where
\begin{equation}
h_n(x,r) = \frac{1}{\Gamma(n)}\int_0^\infty dy \frac{y^{n-1}}
{\sqrt{y^2+x^2}}\frac{1}{e^{\sqrt{y^2+x^2}+r}-1}\;.
\end{equation}
In particular, $h_2(x,r) = -{\rm ln}(1-e^{-(x+r)})$.

The perturbation correction to the partition function
 at the next leading order in the Chern-Simons coupling
$g$ is from the two-loop vacuum diagram given in Fig.~1\footnotemark.
\footnotetext{It is not difficult to check that the
tadpole and so the dumb diagrams have no
contribution, because of the totally anti-symmetric
tensor structure of the Chern-Simons propagator and of the interaction
vertex.}

\unitlength=1.00mm
\linethickness{0.4pt}
\thicklines
\begin{picture}(110.0,33.0)
\put(68.00,18.00){\circle*{2.00}}
\put(82.00,18.00){\circle*{2.00}}
\put(75.00,18.00){\circle{40.00}}
\multiput(68.0,18.0)(2.00,0.00){7}{\line(3,0){1.00}}
\end{picture}
\begin{description}
\item[Fig. 1]
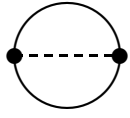
\ \ \ \ The non-vanishing vacuum diagram at order $g^2$.
The real (dashed) lines stand
for the $B_\mu$ ($a_\mu$) propagator.
\end{description}
\vspace{.5cm}

Calculating the two-loop diagram, we obtain
\begin{equation}
{\rm ln}Z_2 = - g^2M\beta V\left(\frac{1}{4M^2}n_0^2+J^2(M,T,\mu)\right)\;,
\end{equation}
and correction to the free energy density
\begin{equation}
{\cal F}_2 
=  \frac{g_r^2M_rT^2}{(4\pi)^2}
\left(\frac{1}{x^2}[g_2(x,-r)
-g_2(x,r)]^2+[x-h_2(x,-r)-h_2(x,r)]^2
\right)\;.
\label{F2}
\end{equation}
In Eq.(\ref{F2}), we have replaced the bare parameters $M$ and $g$
with the renormalized (zero temperature)
parameters $M_r$ (and so the dimensionless parameter $x=M_r/T$)
and $g_r$, whos' definitions to the next leading order will be
given in the next section.

{}From Eq.(\ref{F2}),
we yield correction to particle density at the next leading order,
\begin{eqnarray}
n_1&=&\frac{g_r^2M_r^2}{8\pi^2}
\{\frac{1}{x^2}[g_2(x,-r)-g_2(x,r)][h_2(x,-r)+h_2(x,r)
+x^2h_0(x,-r)
+x^2h_0(x,r)]\nonumber\\
& &~~~~~~~~~~+[x-h_2(x,-r)-h_2(x,r)][g_0(x,-r)-g_0(x,r)]
 \}\;.\label{n1}
\end{eqnarray}
To obtain above, we have used \cite{HW}
\begin{equation}
\frac{\partial}{\partial r} g_{n+1}
=xnh_{n+1} + \frac{x^3}{n}h_{n-1}\;, ~~~~{\rm and}~~~~~
\frac{\partial}{\partial r} h_{n+1}
= \frac{x}{n}g_{n-1}\;.
\end{equation}
Now we see that, when the Chern-Simons interaction is switched on,
the charged particles are attached with the flux and they
are now anyons, the particle density is a function of
the Chern-Simons coupling that characterizes the statistics of the anyons.
If interpreting the anti-particles as ``anti-anyons''
carrying opposite charge with opposite sign of particle
numbers, from Eqs.(\ref{F2}) and (\ref{n1}) we see
 the anyons and anti-anyons interference even
in the ``free'' anyon system.
It would be tempting to check whether the free energy
and other thermodynamical quantities at a particular
Chern-Simons coupling $g^2=\pi/2$ in the present model
turn out to be the ones for the ideal charged fermion
gas, since such an equivalence of the two models at zero temperature
is suggested by the statistics and spin transmutation \cite{CI}.
Unfortunately, it is difficult to check in
perturbation expansion, as all higher order corrections
must be taken into account when the coupling is such strong.

\vskip 0.5truein
\section{Two- and Three-Point Correlations}
\setcounter{equation}{0}
\vspace{3 pt}

In this section, we calculate the two- and three-point correlation functions
to the next leading order and discuss certain
quantities of the system, such as the effective vector mass,
screening lengths, and effective Chern-Simons coupling constant.
We take the chemical potential $\mu=0$, for simplicity.
The corresponding diagrams are depicted in Fig.~2.

We consider first the effective mass of the
$B_\mu$ field. With $\Sigma_{\mu\nu}(p)$ denoting  the self-energy of $B_\mu$,
the inverse two-point correlation function of the $B_\mu$ field is
\begin{equation}
{G_{\mu\nu}}^{-1}(p) = {G^0_{\mu\nu}}^{-1}(p)
-\Sigma_{\mu\nu}(p)\;.
\end{equation}

\unitlength=1.00mm
\linethickness{0.4pt}
\thicklines
\begin{picture}(110.00,32.00)
%
%
\put(9.00,12.00){\line(1,0){22.00}}
\multiput(13.100,13.250)(1.00,2.00){4}{\line(3,0){1.00}}
\put(13.10,12.00){\circle*{2.00}}
\multiput(25.900,13.250)(-1.00,2.00){4}{\line(3,0){1.00}}
\put(26.70,12.00){\circle*{2.00}}
\multiput(18.50,20.90)(2.00,0.00){2}{\line(3,0){1.00}}
\put(6.00, 10.0){\makebox(0,0)[l]{$\mu$}}
\put(32.00, 10.00){\makebox(0,0)[l]{$\nu$}}
\put(17.00, -1.00){\makebox(0,0)[l]{$(a)$}}
\multiput(61.50,12)(-2.00,0.00){3}{\line(3,0){1.00}}
\put(63.00,12.00){\circle*{2.00}}
\put(70.0,12.00){\circle{20.00}}
\put(77.00,12.00){\circle*{2.00}}
\multiput(77.50,12.0)(2.00,0.00){3}{\line(3,0){1.00}}
\put(55.00, 10.0){\makebox(0,0)[l]{$\mu$}}
\put(82.00, 10.00){\makebox(0,0)[l]{$\nu$}}
\put(67.00, -1.00){\makebox(0,0)[l]{$(b)$}}
\put(109.00,12.00){\line(1,0){22.00}}
\multiput(113.100,13.250)(1.00,2.00){4}{\line(3,0){1.00}}
\put(113.10,12.00){\circle*{2.00}}
\multiput(125.900,13.250)(-1.00,2.00){4}{\line(3,0){1.00}}
\put(126.70,12.00){\circle*{2.00}}
\multiput(118.50,20.90)(2.00,0.00){2}{\line(3,0){1.00}}
\put(119.90,12.00){\circle*{2.00}}
\multiput(119.70,11.50)(0.00,-2.00){4}{\line(0,3){1.00}}
\put(106.500, 10.0){\makebox(0,0)[l]{$\mu$}}
\put(131.00, 10.00){\makebox(0,0)[l]{$\nu$}}
\put(121.00, 6.00){\makebox(0,0)[l]{$\lambda$}}
\put(117.00, -1.00){\makebox(0,0)[l]{$(c)$}}
\end{picture}
\begin{description}
\item[Fig. 2]
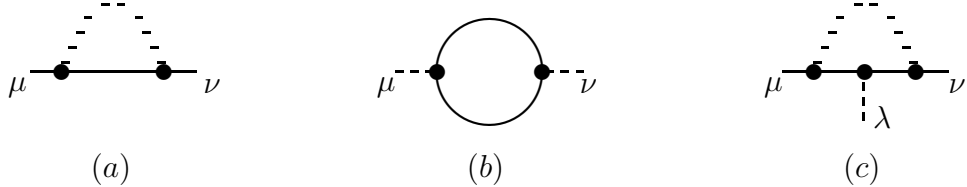
\ \ \ \ One loop diagrams for two- and three-point correlation
functions.
\end{description}
\vspace{.5cm}

To calculate the effective mass,
we set the external momentum-frequence $p=0$, and
consider the self-energy,
from Fig.~2a,
\begin{eqnarray}
\Sigma^{(2)}_{\mu\nu}(0)
&=& g^2T\sum_n\int\frac{d^2q}{(2\pi)^2}
\epsilon_{\mu\sigma\eta}
G^0_{\sigma\lambda}(q)\epsilon_{\lambda\tau\nu}D^0_{\tau\eta}(q)\nonumber\\
&=& 2g^2T\sum_n\int\frac{d^2q}{(2\pi)^2}
\frac{q_\mu q_\nu}{q^2(q^2+M^2)}\;.
\label{Sigma}
\end{eqnarray}
Since the fields are subject to
periodic boundary condition in the imaginary time direction,
and the Lotentz invariance is broken to the spatial rotation
invariance in 2 dimensions,
the longitudinal components of vectors and tensors
are not necessary the same with the transverse ones.
Therefore, the effective mass of the $B_\mu$ field
takes a form \cite{note}
\begin{equation}
M_l(g,T)\delta_{\mu 3}\delta_{\nu 3}
+ M_t(g,T)\delta_{\mu i}\delta_{\nu i}
= {G_{\mu\nu}}^{-1}(p=0)\;.
\label{Mt}
\end{equation}
Calculating Eq.(\ref{Sigma}) and using Eq.(\ref{Mt}),
we obtain the effective masses of $B_\mu$ field
\begin{eqnarray}
M_l(g_r,T) &=&M_r+\frac{g^2_r}{\pi}
M_r\left[\frac{1}{6}
- \frac{1}{x}\left(h_2(x)
- \frac{6}{x^2}[h_4(0)-h_4(x)]\right)\right]
+ {\cal O}(g^4)\;,
\label{MT0}\\
M_t(g_r,T) &=&M_r+\frac{g^2_r}{\pi}M_r
\left[\frac{1}{6}
- \frac{3}{x^3}\left(h_4(0)-h_4(x)\right)\right]
+ {\cal O} (g^4)\;.\label{MT}
\end{eqnarray}
Above, the renormalized (zero temperature) mass $M_r$ is defined by
 $M_r = M-\frac{1}{3\pi^2}g^2\Lambda$, in the regularization
by a naive ultraviolet cut-off $\Lambda$.
In the brackets in Eqs.(\ref{MT0}) and (\ref{MT}),
the bare mass $M$ has been replaced by
the renormalized $M_r$, and the bare (zero temperature)
Chern-Simons coupling $g$ replaced
by the renormalized $g_r$ (its definition will be given later).
These replacements effect on only $g^4$ and higher orders.
Now, we see one of the thermal effects on the vector
mass is a lift of mass degeneracy. The longitudinal effective mass of the
vector is different from the transverse ones.
Moreover, if the chemical potential would be non-vanishing,
it can been seen that the mass matrix
of massive vector field $B_\mu$ develops non-diagonal
components of the form $m(T,\mu)\epsilon_{ij}$.
This results in a further lift of degeneracy of vector mass
in the transverse dimensions when the mass matrix
is diagonalized. The lift of degeneracy in the transverse dimensions
seems to have something to do with the asymmetry of parity,
though we do not discuss this issue further in this paper.

The first terms in the brackets of Eqs.(\ref{MT0}) and (\ref{MT})
are the radiative mass corrections, and
$M_l = M_t =(1+\frac{1}{6\pi} g_r^2)M_r$ at zero temperature.
The finite temperature case is very interesting: due to the
energy exchange with the heat
reservoir, the vector particles (or excitations) ``gain''
weights in their longitudinal dimension, but ``loss'' weights in
transverse ones as $M_l(g_r,T)$ and $M_t(g_r, T)$
are monotonically increasing and
decreasing functions of temperature $T$, respectively, as seen
in Eqs.(\ref{MT0}) and (\ref{MT}).
Moreover, there exists a critical temperature
${T_c}$ at which $M_t(g_r, T) = 0$
\footnotemark. \footnotetext{ An extrapolation of
 (\ref{MT}) to  $T>T_c$
results in a negative self-screening magnetic mass. Since a system with
negative boson mass is not bounded from below, this might imply
one more phase, which is very unstable however.}
$T_c$ is determined by setting Eq.(\ref{MT}) zero.
A typical numerical solution at the next leading order is
\begin{equation}
 T_c \sim 46.3M_r\;,
\label{Tc}
\end{equation}
when $g^2_r = \pi/100$.
Obviously, stronger Chern-Simons interactions correspond to
higher critical temperature.
The phenomenon around $T_c$ seems to be an analogue of the shift
from normal to anomalous
dispersion of vector waves discoved in \cite{BLL}.
We would also speculate that the vector field $B_\mu$
could be both ``electrically'' and ``magnetically'' self-screened
when $T<T_c$, while only the ``electric'' self-screening happened
when $T>T_c$. This might imply a phase transition
between ``conductor'' and ``superconductor'' only when static
electric and magnetic fields would be somehow induced
by the vector field $B_\mu$ itself.

Next, we consider the current-current correlation, and calculate
the screening masses of the plasma.
The electric and magnetic screening masses, ${\cal M}_{el}$
and ${\cal M}_{mag}$, are defined
via the polarization tensor $\Pi_{\mu\nu}(p, g,M_r,T)$ as
\cite{note}:
\begin{equation}
{\cal M}_{el}(g_r,M_r,T)\delta_{\mu 3}
\delta_{\nu 3}
+{\cal M}_{mag}(g_r,M_r,T)
\delta_{\mu i}\delta_{\nu i}
= -\Pi_{\mu\nu}(p=0,g_r,M_r,T)\;.
\label{pi}
\end{equation}
Calculating the one-loop diagram Fig.~2.b with the external $p=0$,
\begin{eqnarray}
 \Pi_{\mu\nu}^{(2)}(0)
&=& g^2T\sum_n\int\frac{d^2q}{(2\pi)^2}
\epsilon_{\mu\sigma\eta}
G^0_{\sigma\lambda}(q)
\epsilon_{\lambda\tau\nu}G^0_{\tau\eta}(q)\nonumber\\
&=& 2g^2T\sum_n\int\frac{d^2q}
{(2\pi)^2}\frac{2q_\mu q_\nu-\delta_{\mu\nu}(q^2+M^2)}{(q^2+M^2)^2}\;,
\label{Pi}
\end{eqnarray}
and using Eq.(\ref{pi}) we
obtain the screening masses
\begin{eqnarray}
 {\cal M}_{el}(g_r,M_r,T)
&=&
\frac{g_r^2}{\pi}M_r\left(\frac{1}{e^x-1}
-\frac{1}{x}{\rm ln}(1-e^{-x})\right)
 + {\cal O}(g^4)\;,\\
 {\cal M}_{mag}(g_r,M_r,T)
&=& 0  + {\cal O}(g^4)\;.
\label{mcs}
\end{eqnarray}
Above we have set the renormalized zero temperature
masses of the gauge field to zero by using
counter terms 
(in the regularization by a large momentum cut-off),
so that the gauge symmetry is respected.
We see now that while the magnetic mass $ {\cal M}_{mag}(g_r,M_r,T)$
vanishes identically to at least the next leading order, electric one
${\cal M}_{el}(g_r,M_r,T)$ is a monotonically increasing function of $T$.
Therefore, like in $QED$ and $QCD$, only the static electric field is
screened by the plasma thermal anyon excitation, and therefore
the plasma thermal excitation acts like a conductor, instead
of superconductor.

In (2+1) dimensions, due to the parity and time reversal asymmetries,
the current-current correlation  may have anti-symmetric  (in the space-time
index) components. Namely, the polarization accepts a decomposition:
\begin{equation}
\Pi_{\mu\nu}(p)
= \epsilon_{\mu\nu\lambda}p_\lambda\Pi_o(p) + {\rm other~terms}\;,
\end{equation}
with $\Pi_o(p=0)$ contributing to the Chern-Simons coefficient.
A simple calculation gives
\begin{equation}
\Pi_o^{(2)}(0) = \frac{g^2_r}{M_r}(1+M_r\frac{\partial}{\partial M_r})J(M_r,T)
=-\frac{g^2_r}{2\pi}\left(1+\frac{1}{e^x-1}
+\frac{1}{x}{\rm ln}(1-e^{-x})\right)\;.
\label{pi0}
\end{equation}
Since the effective Chern-Simons coefficient, together with the
effective Chern-Simons coupling
(to be considered right below), effects on the statistics of anyons,
temperature  dependence of these quantities implies the statistics being
temperature dependent as well.

Finally, we consider the three-point function and determine the effective
Chern-Simons coupling to the next leading
order. The three-point function is defined in the perturbation expansion as
\begin{equation}
\Gamma_{\mu\nu\lambda}(p,k,T)
= \sum_{n=0}g^{2n+1}\Gamma_{\mu\nu\lambda}^{(2n)}(p,k,T)\;.
\end{equation}
Then the effective (finite temperature) coupling constant
is defined through the three-point function at $p=k=0$:
\begin{equation}
\epsilon_{\mu\nu\eta}\delta_{\lambda 0}\delta_{\eta 3}g_1(T)
+\epsilon_{\mu\eta\lambda}
\delta_{\nu 3}\delta_{\eta 0}g_2(T)
=\Gamma_{\mu\nu\lambda}(0,0,T)\;,
\label{gs}
\end{equation}
where $\mu$ and $\nu$ refer to the indexes
of the charged vector fields
$B_\mu$ and $B_\nu^*$ and $\lambda$ for that of the Chern-Simons
gauge field $a_\lambda$.

Calculating the one-loop diagram Fig.~2c,
\begin{eqnarray}
\Gamma^{(2)}_{\mu\nu\lambda}(0,0)
&=&  g^3T\sum_n\int\frac{d^2q}{(2\pi)^2}
\epsilon_{\mu\sigma\eta}
G^0_{\sigma\tau}(q)
\epsilon_{\tau\rho\lambda}G^0_{\rho\eta}(q)\epsilon_{\eta\nu\gamma}
D^0_{\gamma\eta}(q)\;\nonumber\\
&=& g^3T\sum_n\int\frac{d^2q}{(2\pi)^2}
\frac{1}{M}\frac{\epsilon_{\lambda\mu\eta}q_\eta q_\nu -
\epsilon_{\lambda\nu\eta}q_\eta q_\mu }{q^2(q^2+M^2)}\;,
\label{Gs}
\end{eqnarray}
and using Eq.(\ref{gs}),  we obtain the effective coupling
\begin{eqnarray}
g_1(T) &=& g_r-\frac{g_r^3}{\pi}
\left[\frac{1}{6}-\frac{1}{2x}
\left(h_2(x)
-\frac{3}{x^2}[h_4(0) - h_4(x)]\right)\right]
 + {\cal O}(g^5)\;,\label{gt0}\\
g_2(T) &=&  g_r-\frac{g_r^3}{\pi}
\left[\frac{1}{6}-\frac{3}{x^3}[h_4(0)
- h_4(x)]\right]
 + {\cal O}(g^5)\;,
\label{gt}
\end{eqnarray}
where the renormalized zero-temperature coupling
$g_r = g(1 + \frac{g^2\Lambda}{3\pi^2M})$, when
a cut-off is used.
{}From Eqs.(\ref{gt0}) and (\ref{gt}) we see
$g_1=g_2 = g_r(1-\frac{1}{6\pi}g_r^2)$
at absolute zero.  At finite temperatures,
the components of the
effective coupling, $g_1(T)$ and $g_2(T)$, are monotonically
slowly increasing functions of the temperature $T$
(to a reasonably high $T$). In particular, at the critical
temperature $T_c \sim 46.3M_r$, given in (\ref{Tc}), we have
\begin{eqnarray}
g_1(T_c) &\sim& 1.4g_r\;,\\
g_2(T_c) &=& 2g_r\;,
\end{eqnarray}
with $g^2_r = \pi/100$. Therefore, as long as the zero
temperature coupling $g_r$ is small, in a range from zero to a reasonably high
temperature, the effective Chern-Simons coupling is small.

\renewcommand{\theequation}{\thesection.\arabic{equation}}
\vskip 0.5truein
\section{Summary}
\setcounter{equation}{0}
\vspace{3 pt}

We have set up a framework to investigate the thermal behavior of
a relativistic  dynamic anyon theory with a Chern-Simons
field coupling to a massive spin-1 field. We have verified that the
Chern-Simons kinetic term has no contribution to
the free energy and all other thermodynamic quantities,
but the Chern-Simons coupling characterizes these
quantities. Our calculation in perturbation expansion over
a small coupling at finite temperatures supports
the view of point that the sole role of Chern-Simons
interaction is to effect on the statistics of
the matter field coupled.

Our results suggest that, responding to an external
static electric-magnetic field, the anyon system acts like
a conductor, instead of superconductor. However, if
self-induced static electric-magnetic field would
somehow come up (via higher order loop corrections
to the $B_\mu$ self-energy probably),
the system might experience a conductor and superconductor
transition at some critical temperature that depends on
the Chern-Simons coupling and the vector mass. The
observation of finite temperature
correction to the Chern-Simons coefficient
(and the effective
Chern-Simons coupling) might be useful to understanding the
finite temperature quantum Hall effect in a Chern-Simons matter model,
as an effective Chern-Simons
coefficient determines a Hall conductivity.

Since the components of the effective Chern-Simons coupling
are slowly increasing functions of
temperature, as seen in the last section, a perturbation
expansion seems to be reliable in a region from zero
to a reasonably high temperature, as long as the zero temperature
Chern-Simons coupling is small, or as long as
the statistics of anyons is close to that of the vector boson.

Finally, since we expand the anyon system around a free charged
boson with non-vanishing chemical potential,
there seems a concern on Bose-Einstein condensation which
happens in some free (and interaction) boson theories \cite{K};
and if so, how  the (Chern-Simons) interaction that
transmutes bosons into anyons effects on the condensation?
We would like to argue that while this may be an interesting
issue in a model with charged scalar coupled to Chern-Simons field,
the free B-theory studied in this work does not
exhibit a Bose-Einstein condensation.
Let us recall our model involves only first
derivative. This makes it quite apart from
any conventional boson theory. In particular, it does not allow
a zero-momentum term in the Fourier series of $B_\mu$ field,
while such a term is a key for the condensation. To show this,
let us assume the complex field $B_\mu(p=0) = \eta_\mu e^{i\theta_\mu}$,
$(\mu=1,2,3)$. Then the mass term in the action contributes
to the partition function a term ${\rm exp}[M\eta^2/2]$. Accordingly
the chemical potential term contributes
${\rm exp}[i\mu\eta_1\eta_2(e^{-i\theta_1}e^{i\theta_2}
-e^{-i\theta_2}e^{i\theta_1})]$. However,
due to the $U(1)$ global symmetry of lagrangian,
there must be $\theta_1=\theta_2$, so that $\theta_\mu$ does not appear
in the final results. Then the term with chemical
potential vanishes.
To determine $\eta$ $(=|\eta|)$, one sets $\partial
({\rm ln}Z_0)/\partial\eta = 0$, and obtains $\eta M = 0$.
Namely, $\eta = 0$ if $M\neq 0$. This completes the
argument that there is no Bose-Einstein condensation in the B-theory.

The author thanks L. Dolan, C. Itoi, J.Y. Ng, G. Semenoff and Y.-S. Wu
for discussions. This work was supported in part by
the  U.S. DOE under contract No. DE-FG05-85ER-40219.
\vspace{0.50cm}

\baselineskip=18.0truept



\begin{thebibliography}{199}
\bibitem{Green}
H.S. Green, Phys. Rev. {\bf 90}, 270 (1953);
Y. Chnuki and S. Kamefuchi, {\it Quantum Field Theory and Parastatistics},
(Spinger-Verlag, New York, 1982).
\bibitem{any}
E. Merzbacher, Am. J. Phys. {\bf 30}, 237 (1962);
M.G.G. Laidlaw and C.M. DeWitt, Phys. Rev. D {\bf 3}, 1375 (1971);
J. Leinaas and J. Myrheim, Nuovo Cimento {\bf 37}, 1 (1977);
F. Wilczek, Phys. Rev. Lett. {\bf 49}, 957 (1982).
\bibitem{Wil}
A modern review, see F. Wilczek, {\it Fractional Statistics
and Anyon Superconductivity} (World Scientific, Singapore, 1990).
\bibitem{L}
R.B. Laughlin, Phys. Rev. Lett. {\bf 50}, 1395 (1983);
A. Fetter, C. Hanna and R. Laughlin, Phys. Rev. B {\bf 39}, 9679
(1989);
S.M. Girvin and A.H. Macdonald, Phys. Rev. Lett. {\bf 58}
1252 (1987);
S.-C. Zhang, T.H. Hansen and S. Kivelson, {\it ibid}
{\bf 62}, 82 (1989);
D.H. Lee and S.-C. Zhang, {\it ibid} {\bf 66}, 1220 (1991);
N. Read,  {\it ibid}  {\bf 62}, 86 (1989);
J.K. Jain,  {\it ibid} {\bf 63}, 199 (1989);
Phys. Rev. B {\bf 40}, 8079 (1990);
B. Blok and X.-G. Wen,  {\it ibid} {\bf 42}, 8133 (1990);
X.-G. Wen and A. Zee, Nucl. Phys. {\bf B351}, 135 (1990);
X.-G. Wen, Mod. Phys. Lett. {\bf B5}, 39 (1991);
B.I. Halperin, P.A. Lee and N. Read, Phys. Rev. B {\bf 47}, 7312 (1993);
V. Kalmeyer and S.-C. Zhang,  {\it ibid} {\bf 46}, 9889 (1992);
D.H. Lee, S. Kivelson and S.-C. Zhang, Phys. Rev. Lett.
{\bf 68}, 2389 (1992);
X.-G. Wen and Y.-S. Wu, Phys. Rev. Lett. {\bf 70}, 1501 (1993);
W. Chen, M.A.P. Fisher and  Y.-S. Wu, Phys. Rev. B {\bf 48}, 13749 (1993).
\bibitem{L1}
R.B. Laughlin, Phys. Rev. Lett. {\bf 60}, 2677 (1988);
Y.H. Chen, F. Wilczek, E. Witten and B.I. Halperin,
 Int. J. Mod. Phys. {\bf B3}, 1001 (1989); References in \cite{Wil}.
\bibitem{LRF}
A. Lopez, A.G. Rojo, and E. Fradkin, {\it Chern-Simons Theory
of the Aniotropic Quantum Heisenberg Antiferromagnet on a Square Lattice},
 Phys. Rev. B (1994). 
\bibitem{tem}
A. Niemi, Nucl. Phys. {\bf B251}, 155 (1985);
R.D. Pisarski, Phys. Rev. D {\bf 35}, 664 (1987);
K. Baju, A. Das, and P. Panigrahi,
{\it ibid} {\bf 36}, 3725 (1987);
G.W. Semenoff, P. Sodano and Y.-S. Wu, Phys. Rev. Lett. {\bf 62}, 715 (1989);
E. Poppitz, Phys. Lett. {\bf B252}, 417 (1990);
M. Burgess, Phys. Rev. D {\bf 44}, 2552 (1990);
L. Moriconi,  {\it ibid}, R2950 (1990);
J.D. Lykken, J. Sonnenschein, and N. Weiss, Int. J. Mod. Phys. {\bf A6},
1335, (1991);  Y.-C. Kao, Mod. Phys. Lett. {\bf A6}, 3261 (1991);
A. Das and S. Panda, J. Phys. {\bf A25}, L245 (1992);
I.J.R. Atichinson, C.D. Fosco and J.A. Zuk,
Phys. Rev. D {\bf 48}, 5895 (1993);
G. Gat and R. Ray, preprint (1994).
N. Bralic, D. Cabra, and F.A. Schaposnik,
preprint (1994).
\bibitem{P}
A.M. Polyakov, Mod. Phys. Lett. {\bf A3}, 325 (1988);
C.-H. Tze, Int. J. Mod. Phys. {\bf A3}, 1959 (1988);
S. Iso, C. Itoi and H. Mukaida, Nucl. Phys. {\bf B346}, 293 (1990);
N. Shaju, R. Shankar and M. Sivakumar, Mod. Phys. Lett. {\bf A5},
593 (1990).
\bibitem{CI}
W. Chen and C. Itoi,
Phys. Rev. Lett. {\bf 72},  2527 (1994);
Nucl. Phys. {\bf B435} [FS], 690 (1995).
\bibitem{K}
see, for example, D.J. Gross, R.D. Pisarski and L.G. Yaffe,
Rev. Mod. Phys. {\bf 53}, 43 (1981);
J.I. Kapusta, {\it Finite-Temperature
Field Theory}, (Cambridge Univ. Press, New York, 1989).
\bibitem{TPN}
P.K. Townsend, K. Pilch and P. van Nieuwenhuizen,
Phys. Lett. {\bf 126B}, 38 (1984).
\bibitem{DJ}
S. Deser and R. Jackiw, Phys. Lett. {\bf 139B}, 38 (1984).
\bibitem{IIM}
S. Iso, C. Itoi, and H. Mukaida, Phys. Lett. {\bf 236B}, 287 (1990).
\bibitem{JN}
R. Jackiw and V.P. Nair, Phys. Rev. D {\bf 43}, 1933 (1991).
\bibitem{WZ}
X.G. Wen and A. Zee, Phys. Rev. Lett. {\bf 62}, 2873 (1989).
\bibitem{BLL}
M. Burgess, J.M. Leinaas and O.M. Lovvik, Phys. Rev. B {\bf 48}, 12912 (1993).
\bibitem{Huang}
see, for example, K. Huang, {\it Statistical Mechanics}
(Wiley, New York, 1963).
\bibitem{Wein}
 L. Dolan and R. Jackiw,
Phys. Rev. D {\bf 9}, 3320 (1973);
S. Weinberg,
Phys. Rev. D {\bf 9}, 3357 (1973).
\bibitem{HW}
H.E. Haber and H.A. Weldon, J. Math. Phys. {\bf 23}(10), 1852 (1982).
\bibitem{note}
Here, as normal, the effective masses are defined at
$p_3 = 0$ and ${\bf p}\rightarrow 0$. It is equivalent to take
$p=0$ in the present case. This is because the expansions of
$\Sigma(0,{\bf p})$ and $\Pi(0, {\bf p})$ at ${\bf p} = 0$ are
infrerad convergent, due to the non-zero mass parameter $M$.
\end{thebibliography}
\end{document}